# Electrically-Tunable Stochasticity for Spin-based Neuromorphic Circuits: Self-Adjusting to Variation

Hossein Pourmeidani,[1] Punyashloka Debashis,[2] Zhihong Chen,[2] Ronald F. DeMara,[1] and Ramtin Zand[3]

[1]Department of Electrical and Computer Engineering, University of Central Florida, Orlando, FL 32816, USA
[2]School of Electrical and Computer Engineering, Purdue University, West Lafayette, IN 47907, USA
[3]Department of Computer Science and Engineering, University of South Carolina, Columbia, SC 29208, USA

*Abstract*—Energy-efficient methods are addressed for leveraging low energy barrier nanomagnetic devices within neuromorphic architectures. Using a Magnetoresistive Random Access Memory (MRAM) probabilistic device (p-bit) as the basis of neuronal structures in Deep Belief Networks (DBNs), the impact of reducing the Magnetic Tunnel Junction's (MTJ's) energy barrier is assessed and optimized for the resulting stochasticity present in the learning system. This can mitigate the process variation sensitivity of stochastic DBNs which encounter a sharp drop-off when energy barriers exceed near-zero *kT*. As evaluated for the MNIST dataset for energy barriers at near-zero *kT* to 2.0 *kT* in increments of 0.5 *kT*, it is shown that the stability factor changes by 5 orders of magnitude. The self-compensating circuit developed herein provides a compact, and low complexity approach to mitigating process variation impacts towards practical implementation and fabrication.

## I. INTRODUCTION AND BACKGROUND

Stochastic computing has been studied as the foundation of brain-inspired computing frameworks. However, hardware implementation of stochastic computing models have received less consideration despite the use of stochasticity being widespread in the computational neuroscience field. In many cases, the underlying CMOS hardware used to imitate neuronal functionality has been deterministic and would thus require area-consuming random number generators to realize stochastic behavior. On the contrary, post-CMOS technologies such as spintronic devices exhibit inherent randomness during their switching processes. Recently, the stochasticity behavior of these single-bit spin-based hardware units has been used to realize compact and effective neuronal hardware in various learning systems [1]-[4].

A recently-proposed building block based on embedded Magnetoresistive Random Access Memory (MRAM) technology is the probabilistic p-bit device [5]. This hardware utilizes a 2-terminal magnetic tunnel junction (MTJ) having two feasible resistive levels depending on the orientation of its ferromagnetic (FM) layers. The FM layers involve a fixed layer that contains a fixed magnetic orientation and a free layer with a magnetization orientation that can be switched. In conventional MRAM cells, the free layer is fabricated with a thermally-stable nanomagnet with a large energy barrier in accordance with the thermal energy, *kT*. Consequently, the fixed layer operates as a single-bit non-volatile storage element whereas providing a thresholding behavior suitable for neural network applications. In recent years, several studies theoretically and experimentally have investigated the usage of thermally-unstable MTJs with near-zero energy barrier based on superparamagnetic materials to realize probabilistic neuromorphic paradigms using functional spintronic devices [2]-[4]. For instance, [6][7] focus on leveraging p-bit devices in deep belief network (DBN) architectures. In this paper, we seek to examine the effects of process variation (PV) on the energy barrier of the p-bit devices and their consequent impacts on the accuracy and energy consumption of a representative neuromorphic DBN architecture. Finally, we propose a method based on temporal redundancy, as well as a circuit-level approach to address the PV-induced performance challenges of p-bit based DBNs.

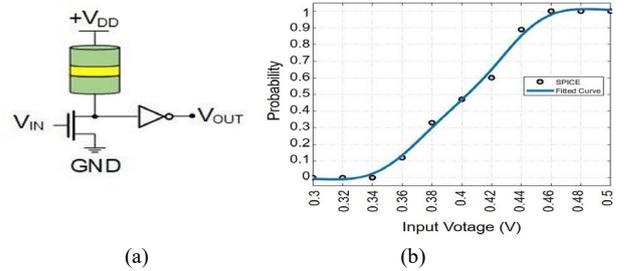

Figure 1: (a) Embedded MRAM-based neuron (p-bit) [5], (b) The output probability of p-bit being in state "1" vs. its input voltage over for 14nm PTM with $V_{DD}$=0.8V nominal voltage.

## II. P-BIT AS A STOCHASTIC NEURON

Camsari et al. [5] proposes a thermally-unstable MRAM device with a low energy-barrier nanomagnet ($E_b \ll 40\ kT$). The MTJ resistance of this device arbitrarily fluctuates between the two possible resistive states. As a consequence, the output voltage at the drain of the NMOS transistor is fluctuating. Moreover, a CMOS inverter which is modulated by the input voltage will amplify such voltage deviation from the threshold voltage in order to produce a stochastic sigmoidal output. As the drain-source resistance ($r_{ds}$) is increased by reducing the input voltage ($V_{IN}$), the voltage at the drain of the NMOS transistor is raised to $V_{DD}$. On the other hand, it is shorted to ground by decreasing the $r_{ds}$ through an increase of $V_{IN}$. Such device operation is formulated considering the MTJ conductance:

$$G_{MTJ} = G_0 \left[ 1 + m_z \frac{TMR}{(2 + TMR)} \right] \quad (1)$$

where $m_z$ is the free layer magnetization, $G_0$ denotes the average MTJ conductance, $(G_P + G_{AP})/2$, and TMR represents the tunneling magnetoresistance ratio. The drain voltage can be written as:

$$V_{DRAIN}/V_{DD} = \frac{(2 + TMR) + TMR\ m_z}{(2 + TMR)(1 + \beta) + TMR\ m_z} \quad (2)$$

where $\beta$ is the ratio of the transistor conductance ($G_T$) to the average MTJ conductance ($G_0$). Figure 1 shows the relation between the probability of p-bit's output being in state "1" and $V_{IN}$. A close observation shows that $V_{IN} = \frac{V_{DD}}{2} = 400$ mV

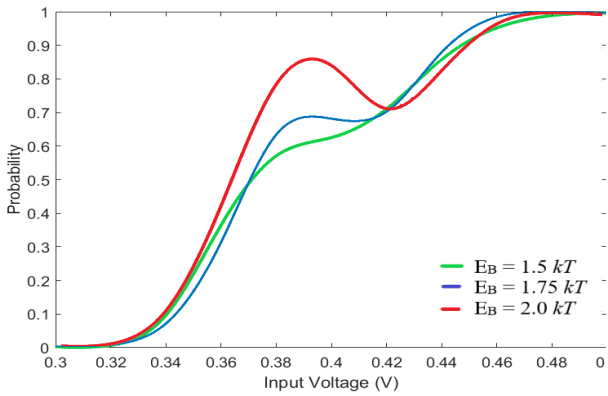

Figure 2: Output probability of MRAM-based neuron for $E_B = 1.5\ kT$, $E_B = 1.75\ kT$, and $E_B = 2.0\ kT$.

leads to approximately 50% probability of p-bit's output being in state "1".

### III. EFFECTS OF PROCESS VARIATION ON P-BIT

The p-bit device is not entirely tolerant of defects and device-to-device variations even though is more error resilient than strictly digital computing devices [8]. The statistical distribution of the magnetization fluctuations, such as the power spectral density become affected by the presence of both localized and delocalized structural defects and moderate variations for the barrier height of the nanomagnet which is caused by small size variations [9]. It is investigated that the power spectral density is relatively insensitive to the presence of small localized defects and moderate barrier height change. Nevertheless, the power spectral density is substantially affected by delocalized defects such as thickness variations over a significant fraction of the nanomagnet [10]. Delocalized defects can considerably change the fluctuation rate of the magnetization in low barrier nanomagnets. This will affect applications in p-bit-based neurons for neuromorphic architectures because the fluctuation rate is essential for stochastic computing applications.

The near-zero energy barrier in p-bit devices is achievablee by reducing the total magnetic moment through decreasing volume (V) and/or manage a small anisotropy field ($H_K$) [11], according to the below relation:

$$E_B = \frac{1}{2} H_K M_S V = \frac{1}{2} H_K M_s (\pi (d/2)^2 t_f) \quad (3)$$

where $d$ and $t_f$ are the diameter and thickness of the MTJ's free layer. Due to the variations in the fabrication process of low energy barrier nanomagnets, p-bits may exhibit different energy barriers [12]. Based on (4), variations in MTJ's anisoptropy field ($\sigma H_K$) and nanomagnet diameter ($\sigma d$) cause linear and quadratic variations in energy barrier ($\sigma E_B$), respectively. As described in previous section, the near-zero energy barrier free layer will fluctuate arbitrarily between the parallel and anti-parallel magnetic states. The magnetization dwell time in the parallel and anti-parallel states creates a distribution which confirms that the nanomagnet fluctuates stochastically. By switching the magnetization direction of the free layer between parallel and anti-parallel states, a sigmoidal distribution is observed over a sequence of samples. These state transitions are instigated by thermal energy which is adequate to randomly fluctuate when using a sufficiently small energy barrier. The fluctuation speed of a nanomagnet can be obtained from the average dwell time in parallel and anti-parallel states $\tau_P$ and $\tau_{AP}$ as follows [13], $\tau^{-1} = \tau_P^{-1} + \tau_{AP}^{-1}$, and, this time scale is related to the energy barrier ($E_B$) of the nanomagnet through $\tau = \tau_0 \times \exp(E_B/K_B T)$. Thus, the fluctuation speed of nanomagnet can be increased or decreased by reducing or increasing the energy barrier, respectively, which will impact the probabilistic behavior of the p-bit devices as described in Section V.

The defects caused by the fabrication imperfections are required to be addressed for neuromorphic applications using p-bit based neurons such as DBNs due to their significant impact on their performance and accuracy. Generally, these challenges raised by variations can be addressed by two approaches. Firstly, a fabrication-oriented approach aims to refine materials and production processes. Alternatively, a post-fabrication mechanism is proposed herein which leverages temporal redundancy as well as a circuit-level mechanism to address the aforementioned PV-imposed challenges. Moreover, a sensitivity-analysis will be conducted to inform the production process with the acceptable range and tolerances for critical parameters impacting the energy-barrier and resulting stochasticity of the p-bit device.

### IV. PROPOSED VARIATION-IMMUNE P-BIT IMPLEMENTATION

Herein, the impact of energy barrier variation is assessed by using a random distribution of parameters for several ranges from near-zero kT to 2.0 kT. The higher energy barrier of 1.5 kT, 1.75 kT, and 2.0 kT are realized by increasing the small anisotropy field ($H_K$). As mentioned, increasing the energy barrier decreases the probabilistic fluctuation speed of the nanomagnet in p-bit devices, which means if we do not change the sampling time of the p-bit's output of probabilistic sigmoidal activation function shown in Figure 1 will be distorted as shown in Figure 2. The results obtained by MATLAB simulation, depicted in Figure 3, show that while the energy barriers less than or equal to 1.5 kT yield an recognition error of approximately 5% (i.e., accuracy rate ~95%) for MNIST hand-written digit recognition application, the error rate will be drastically increased to an unacceptable value of ~90% (i.e. accuracy rate ~10%) for the energy barriers more than 1.75 kT on a 784×200×10 DBN which is trained by 60,000 training images. Thus, the process variation sensitivity of DBNs utilizing low energy barrier MTJs are seen to encounter a sharp "knee effect" drop-off when energy barriers exceed 1.75 kT as illustrated in Figure 3.

#### A. P-bit with temporal redundancy

The increase in the energy barrier of p-bit results in a decrease in the fluctuation speed of its nanomagnet. It means that given sufficiently long sampling time, the p-bit's output voltage can realize its probabilistic sigmoidal behavior without notable distortions. To verify the effect of increasing the sampling window period ($\tau_S$) of p-bit's output to address the energy-barrier variation issues, we have examined p-bits with four different energy barriers: $0.5\ kT$, $1\ kT$, $1.5\ kT$, and $2\ kT$. Figure 4 shows an experiment conducted in SPICE circuit

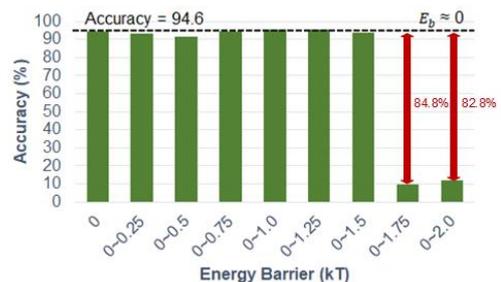

Figure 3: Effects of neuron's energy barriers on the DBN accuracy.

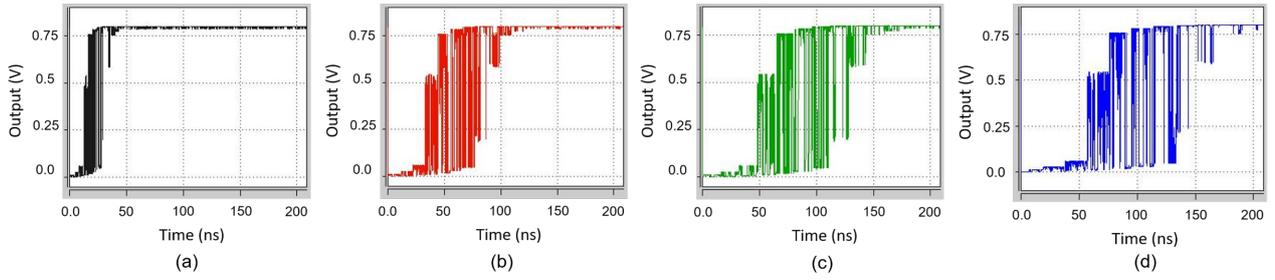

Figure 4: Output of MRAM-based neuron vs. time for different energy barriers (a) $E_B = 0.5\ kT$, (b) $E_B = 1.0\ kT$, (c) $E_B = 1.5\ kT$, and (c) $E_B = 2.0\ kT$.

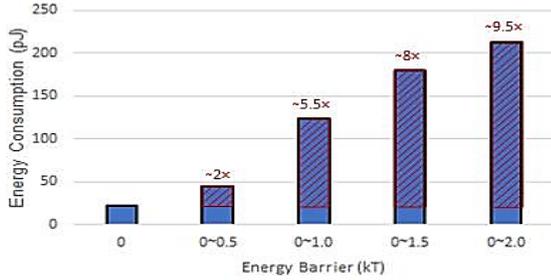

Figure 5: Effect of temporal redundancy on the energy consumption of a variation-tolerant p-bit based DBN with a 784×200×10 topology.

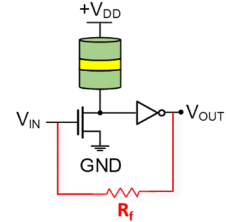

Figure 6: In-circuit adaptation of p-bit using resistive feedback.

$$f_0 = (\tau_0 \times \exp(E_B/K_B T))^{-1} \qquad (4)$$

simulator, in which the input voltage of the p-bit neurons with different energy barriers is incrementally increased from 0.3V to 0.5V (i.e. the active region of the p-bits probabilistic sigmoidal activation function) with 20mV steps. In every step the input voltage remains fixed for $\tau_S$ period and the output voltage is monitored. It is shown that in order to achieve the sigmoidal output required to be realized by p-bit based neurons with $0.5\ kT$, $1\ kT$, $1.5\ kT$, and $2\ kT$, the minimum $\tau_S$ should be tuned to 4ns, 11ns, 16ns, and 19ns, respectively, while the sampling window period for a p-bit with near-zero energy barrier is 2 ns. These results are obtained using the SWEEP function provided by HSPICE circuit simulator.

To assess differences in energy consumption of DBNs under different energy barriers, we examined a 784×200×10 DBN circuit implemented by the PIN-Sim framework [7] for MNIST digit recognition application. We have used p-bits models with the maximum energy-barrier variations ranging from ~0 $kT$ to 2.0 $kT$ with 0.5 $kT$ steps. Figure 5 illustrates the energy consumptions of DBNs with various levels of energy barrier variation tolerance using the proposed temporal redundancy mechanism. The energy that is consumed in DBN with ~2kT energy barrier variation tolerance is approximately 10-fold greater than variation-less DBNs utilizing p-bits with near-zero energy barrier. The variations are applied via PIN-Sim tool by using a randomly generated energy barrier value between $0kT$ and a maximum energy barrier variation defined by user. Thus, in terms of energy consumption, a "knee effect" point for the energy barrier is seen to be around 0.5 kT for our DBN. This knee effect factor can be alleviated in practice by configuring a feedback mechanism to increase the fluctuation rate of the nanomagnet, as described in the following Section.

### B. P-bit with feedback

In this Section, we demonstrate a p-bit neuron circuit, in which the fluctuation rate of its nanomagnet can be tuned using an electrical feedback. In this neuron, the average fluctuation frequency ($f_0$) is determined by the energy barrier of the nanomagnet through the following equation [13]:

Herein, the output of the p-bit device is amplified and fed back to the NMOS transistor, thus the magnetization fluctuation becomes faster, depending on the polarity and strength of the feedback, as a modulation method to compensate towards optimal levels of thermal noise. An implementation of the feedback configuration is illustrated in Figure 6. In this case, the drain of the NMOS transistor tracks the magnetization direction of the free layer of the MTJ. The inverter at the output of the device naturally generates the inverse voltage, hence realizing a feedback compensation mechanism. The feedback can be controlled by changing the value of the resistor $R_f$, which changes the feedback current flowing through the NMOS transistor. Figure 7 (a) shows the output of p-bit with an energy barrier of $E_B \approx 1.5$ kT and a feedback with $R_f$ = 100 KΩ, while in the no feedback case ($R_f$ = infinity), the nanomagnet of device fluctuates extremely slower as shown in Figure 7 (b). Employing the feedback resistor decreases the fluctuation time scale, $\tau$, by approximately 5 times, which reduces the need for temporal redundancy, in consequent of which the energy consumption in the variation-tolerance p-bit neuron will be decreased.

A similar experiment involving feedback of the p-bit output to its input was performed in a current controlled device scheme [14]. The effect of feedback on frequency tunability can be understood by considering the change to the energy landscape of the nanomagnet. In the feedback configuration, when magnetization is in the "P" state, the device output feeds back a negative current to its input, thus tilting the energy barrier in favor of the "AP" state, i.e, the barrier that needs to be overcome to transition from the "P" to the "AP" state becomes smaller than the barrier for the reverse transition. Similarly, when the magnetization is in the "AP" state, the barrier for transitioning from the "AP" to the "P" state is smaller than the barrier for the reverse transition. So, the energy landscape is dynamically modified in a way such that the energy barrier appears to be lower to transition from the occupied state to the other state. This effect increases the fluctuation frequency of the device output, expressed as:

$$f_0 = (\tau_0 \times \exp(E_{B,eff}/K_B T))^{-1} \quad (5)$$

where the effective energy barrier ($E_{B,eff}$) is given by:

$$E_{B,eff} = E_B(1 \pm I_{feedback}/I_C) \quad (6)$$

where $E_B$ is the intrinsic energy barrier of the nanomagnet given in equation (3), $I_{feedback}$ is the feedback current and $I_C$ is the critical current for magnetization switching at zero temperature. $I_{feedback}$ can be replaced by $V_{DD}R_f$ from analyzing the circuit configuration, considering that the NMOS transistor resistance is much smaller than $R_f$ (which can be realized by choosing a large enough $R_f$). Next, by defining $V_{DD}I_C$ as $R_0$, we get the following expression for the effective energy barrier of the magnet:

$$E_{B,eff} = E_B(1 \pm R_0/R_f) \quad (7)$$

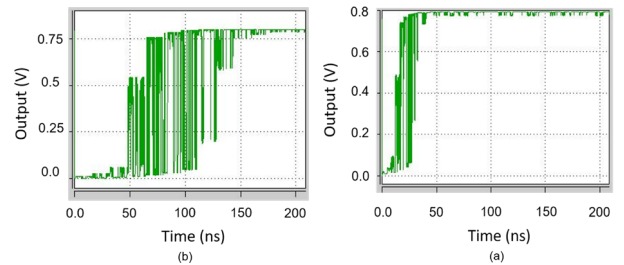

Figure 7: The output fluctuations of the device for $E_B = 1.5\ kT$, (a) with 100 KΩ feedback resistor, (b) without the feedback resistor.

Equations (5) and (7) elaborate that the fluctuation frequency of the p-bit can be controlled by changing the feedback resistor, as also demonstrated in the co-author's experiment [14]. The above analysis generally holds true for the device presented in this paper. The circuit simulation results exhibit that maximum variations of 0.5 $kT$, 1 $kT$, 1.5 $kT$ and 2 $kT$ can be compensated using $R_f$ with $30K\Omega$, $60K\Omega$, $100K\Omega$, $120K\Omega$ resistances, respectively. This is realized with only ~12% energy overhead, which is 25.1 pJ for p-bit with $120K\Omega$ feedback resistor compared to 22.4 pJ for p-bit without feedback.

## V. Conclusion

Herein, we investigated two approaches to mitigate the effects of process variation on the energy barrier of the p-bit based neurons, and their consequent impact on the performance and accuracy of DBNs using p-bit devices as probabilistic sigmoidal neurons. In the first approach, it was shown that an increase in the energy barrier leads to decreased fluctuation speed in the magnetization direction of the p-bit's nanomagnet. Thus, to observe the desired probabilistic sigmoidal behavior in the p-bit based neuron a temporal redundancy is required to be added to the sampling time of the p-bits output to provide an operating interval sufficient for probabilistic fluctuations. While the temporal redundancy has shown to be an efficient mechanism, it was examined that it can lead to approximately 10-fold higher energy consumption in a 784×200×10 DBN which can tolerate maximum 2 $kT$ of energy barrier variations compared to a variation-less DBN. The second variation tolerance mechanism involved implementing p-bit with a negative self-feedback, which significantly increases the probabilistic fluctuation speed of the free layer. In this case, the drain of the NMOS transistor in the p-bit device tracks the magnetization direction of the free layer of the MTJ, and the inverter at the output of the device generates the inverse voltage, hence realizing a negative feedback effect which compensates the variation impacts with only ~10% energy consumption overheads.

## Acknowledgment

This work was supported in part by the Center for Probabilistic Spin Logic for Low-Energy Boolean and Non-Boolean Computing (CAPSL), one of the Nanoelectronic Computing Research (nCORE) Centers as task 2759.006, a Semiconductor Research Corporation (SRC) program sponsored by the NSF through CCF 1739635.

## References

[1] W. H. Choi et al., "A Magnetic Tunnel Junction based True Random Number Generator with conditional perturb and real-time output probability tracking," *Tech. Dig. - Int. Electron Devices Meet. IEDM*, vol. 2015-February, no. February, pp. 12.5.1-12.5.4, 2015.

[2] P. Debashis, R. Faria, K. Y. Camsari, J. Appenzeller, S. Datta, and Z. Chen, "Experimental demonstration of nanomagnet networks as hardware for Ising computing," *Tech. Dig. - Int. Electron Devices Meet. IEDM*, pp. 34.3.1-34.3.4, 2017.

[3] A. Sengupta, M. Parsa, B. Han, and K. Roy, "Probabilistic Deep Spiking Neural Systems Enabled by Magnetic Tunnel Junction," *IEEE Trans. Electron Devices*, vol. 63, no. 7, pp. 2963–2970, 2016.

[4] V. Ostwal, P. Debashis, R. Faria, Z. Chen, and J. Appenzeller, "Spin-torque devices with hard axis initialization as Stochastic Binary Neurons," *Sci. Rep.*, vol. 8, no. 1, pp. 1–8, 2018.

[5] K. Y. Camsari, S. Salahuddin, and S. Datta, "Implementing p-bits with Embedded MTJ," IEEE Electron Device Lett., vol. 38, no. 12, pp. 1767–1770, 2017.

[6] H. Pourmeidani, S. Sheikhfaal, R. Zand, and R. F. DeMara, "Probabilistic Interpolation Recoder for Energy-Error-Product Efficient DBNs with p-bit Devices," *IEEE Transactions on Emerging Topics in Computing*, 2020.

[7] R. Zand, K. Y. Camsari, S. Datta, and R. F. DeMara, "Composable probabilistic inference networks using MRAM-based stochastic neurons," *ACM J. Emerg. Technol. Comput. Syst.*, vol. 15, no. 2, 2019.

[8] J. L. Drobitch and S. Bandyopadhyay, "Reliability and Scalability of p-Bits Implemented With Low Energy Barrier Nanomagnets," *IEEE Magnetics Letters*, vol. 10, pp. 1–4, 2019.

[9] M. A. Abeed and S. Bandyopadhyay, "Sensitivity of the power spectra of thermal magnetization fluctuatuions in low barrier nanomagnets proposed for stochadtic computing to in-plane barrier height variations and structural defects," SPIN, 2050001, *World Scientific Publishing Company*, 2019.

[10] J. Kaiser, A. Rustagi, K. Y. Camsari, J. Z. Sun, S. Datta, and P. Upadhyaya, "Subnanosecond Fluctuations in Low-Barrier Nanomagnets," Phys. Rev. Applied 12, 054056, 2019.

[11] K. Y. Camsari, B. M. Sutton, and S. Datta, "P-bits for probabilistic spin logic," *Appl. Phys. Rev.*, vol. 6, no. 1, 2019.

[12] J. L. Drobitch and S. Bandyopadhyay, "Spin electronics robustness and scalability of p-bits implemented with low energy barrier nanomagnets," pp. 1–5, 2019.

[13] P. Debashis, R. Faria, K. Y. Camsari, S. Datta, and Z. Chen. "Correlated fluctuations in spin orbit torque coupled perpendicular nanomagnets." Physical Review B 101, no. 9 (2020): 094405.

[14] P. Debashis and Z. Chen. "Electrical annealing and stochastic resonance in low barrier perpendicular nanomagnets for oscillatory neural networks." In 2019 77th Device Research Conference (DRC), pp. 85-86. IEEE, 2019.